\documentclass[fleqn,3p,onecolumn,12pt,nofootinbib,showkeys,showpacs,
               preprint,eqsecnum]{revtex4-1}

\usepackage[small,justification=centerlast]{caption}
\usepackage{graphicx}
\usepackage{bm}
\usepackage{mathrsfs}
\usepackage[latin1]{inputenc}
\usepackage{array}
\usepackage{psfrag}
\usepackage{dsfont}
\usepackage{epsfig}
\usepackage{titletoc}
\usepackage{float}   
\usepackage{wrapfig} 
\usepackage{amsmath}
\usepackage[latin1]{inputenc}
\usepackage{amsmath}
\usepackage{amssymb}
\usepackage{mathrsfs}
\usepackage{array}
\usepackage{graphicx}
\usepackage{psfrag}
\usepackage{dsfont}
\usepackage{epsfig}
\usepackage{cancel}
\usepackage{upgreek}
\usepackage{subfig}
\usepackage{tabularx}
\usepackage{color}
\usepackage{units}

\usepackage{hhline}

\usepackage{textfit} 

\usepackage{hyperref}

\begin{document}

	


\noindent  J. Math. Phys. \textbf{55}, 112503 (2014)
\hfill  arXiv:1404.2901\newline\vspace*{2mm}

\title{Comparison of spacetime defects which are homeomorphic but not diffeomorphic}
\author{F.R.~Klinkhamer}
\email{frans.klinkhamer@kit.edu}
\author{F. Sorba}
\affiliation{Institute for Theoretical Physics \\ Karlsruhe Institute of Technology \\
76128 Karlsruhe, Germany\\}

\begin{abstract}
Certain remnants of a quantum spacetime foam can be modeled by a distribution of defects embedded in a flat classical spacetime. The presence of such spacetime defects affects the propagation of elementary particles. In this article, we show explicitly that both topology and differential structure of the defects are important for the particle motion. Specifically, we consider three types of spacetime defects which are described by the same topological manifold $\mathbb{R}\times\big(\mathbb{R}P^3-\{\text{point}\}\big)$ but which are not diffeomorphic to each other. We investigate the propagation of a massless scalar field over the three different manifolds and find different solutions of the \mbox{Klein--Gordon} equation.
\end{abstract}

\maketitle
\newpage
\begingroup
\hypersetup{pdfborder = {0 0 0}}
\endgroup
\newpage

\section{Introduction}\label{sec:introduction}

Spacetime foam may be one of the features of the
quantum theory of gravity: at microscopic length scales comparable
to the Planck length, spacetime is
affected by quantum fluctuations of its geometry and topology~\cite{Wheeler1955,Wheeler1957}.
It is interesting, then, to investigate how these fluctuations influence the propagation of particles in the emerging classical spacetime.
Many different models have been proposed over the years to
represent the effects of spacetime foam (see, e.g.,
Refs.~\cite{AmelinoCamelia2008,DowkerHensonSorkin2010,Hossenfelder2013a}
and references therein). Here, we are particularly interested
in the approach used in Ref.~\cite{BernadotteKlinkhamer2006}.

In that work, the possible remnants of the spacetime foam are described by a Swiss-cheese-type classical spacetime manifold, where balls of space are removed from a spatial slice of Minkowski spacetime. The holes of this manifold have antipodal points on their boundaries identified and we refer to the resulting structures by the name of ``spacetime defects.'' In this case, space around a defect is simply described by spherical coordinates and the topological structure of the defect is implemented by additional boundary conditions on the matter fields. But a system of spacetime coordinates that automatically describes the structure
of the defect has not been supplied in Ref.~\cite{BernadotteKlinkhamer2006}.

A first attempt to solve this problem has been provided in Ref.~\cite{Schwarz2010},
where it has been shown that it is indeed possible to introduce such a system of
spacetime coordinates. However, the particular coordinates discussed in that work
were used to define a manifold whose metric is not smooth at the 
defect surface and is not a vacuum solution  
of Einstein's gravitational field equations.

An improved result has been obtained recently in Ref.~\cite{KlinkhamerRahmede2013}. In that work, a different set of spacetime coordinates has been introduced,
which allows us to define a defect manifold whose metric is smooth everywhere. The form of the metric has been derived as a vacuum solution of the Einstein field equations.

The three types of spacetime defects described in Refs.~\cite{BernadotteKlinkhamer2006,Schwarz2010,KlinkhamerRahmede2013} have the same topological structure. The three manifolds are indeed related by homeomorphisms, but they are not diffeomorphic to each other~\cite{Klinkhamer2013,Klinkhamer2013-review}. Here, we show that this is a physically relevant distinction and that the three manifolds really describe different types of spacetime defects with different observable characteristics. To do so, we compare the solutions of the massless Klein--Gordon equation for the three cases.

At this moment, a few remarks may help to place our spacetime defects
in context.
First, of the three spacetime defects considered in this paper only one is a
solution of the vacuum Einstein equations. 
Second, that particular vacuum solution
(with a mass-type parameter $\ell>0$) appears asymptotically in a
finite-energy matter solution with the same topology~\cite{Klinkhamer2014}.
This matter solution with a Skyrmion field has, however,
only been obtained numerically
and the present paper focuses instead on the vacuum solution which is
known analytically.
Third, the vacuum defect solution is reminiscent of the so-called
$\mathbb{R}P^{3}$--geon solution~\cite{Friedman-etal1993}
(further references and a brief review can be found in~\cite{Louko2010}).
The $\mathbb{R}P^{3}$--geon can be interpreted as
a non-static defect with length scale $b$ ranging from
$0$ (at the initial and final curvature singularities)
to $2GM/c^2$ (halfway between the initial and final singularities), whereas the
defect vacuum solution is static and has constant length scale $b > 2GM/c^2$.
Fourth, the metric of the defect vacuum solution is
non-Lorentzian, i.e., the standard elementary flatness condition
does not apply everywhere~\cite{Klinkhamer2013,Klinkhamer2013-review};
see below for details. The motivation of the present paper is to
to better understand this non-Lorentzian metric
(independent of coordinate issues) by studying the
Klein--Gordon equation. As such, the scope of the present paper is limited.

The structure of the paper is as follows.
In Sec. \ref{sec:framework}, we introduce the three types of defects.
In Sec. \ref{sec:scalar_solutions},
we discuss and compare the solutions of the Klein--Gordon equation for the
three different cases.
In Sec. \ref{sec:conclusion}, we give a summary of the results.
In App. \ref{app:radial_solution_0}, we provide a detailed derivation of the solutions near the defect surface for the smooth manifold from
Ref.~\cite{KlinkhamerRahmede2013} and compare to the case of standard Minkowski spacetime.

\section{Framework}
\label{sec:framework}

\subsection{Types of manifolds}
\label{subsec:Types-of-manifolds}

In this section, we introduce the different spacetime defects
to be examined later on. The defect length scale is
denoted by $b$. We start with the simplest case, where the defect is
obtained from Minkowski spacetime by surgery~\cite{BernadotteKlinkhamer2006}
and the resulting spacetime is denoted
$\widehat{M}_{b}$, which is a topological manifold with a
differential structure inherited from Minkowski spacetime.
We, then, consider the defect described by the
manifold from Ref.~\cite{Schwarz2010}, referred to as $\widetilde{\mathcal{M}}_{b}$,
which is a differentiable manifold but not a smooth
Lorentzian (pseudo-Riemannian) manifold.
Finally, we introduce the manifold from Ref.~\cite{KlinkhamerRahmede2013},
denoted $\mathcal{M}_{b}$, which would be a smooth Lorentzian manifold,
were it not for the fact that at certain points the metric is degenerate and
the standard elementary-flatness property does not hold
(details will be given in Sec.~\ref{subsec:smooth_defect}).

Let us clarify the distinctions between these different types
of manifold.
With a ``topological manifold'' $M_T$ is meant a topological space as defined   
in Ref.~\cite{Nakahara2003}
(that is, a set $X$ equipped with a topology $\mathcal{T}$),
which is locally homeomorphic to $\mathbb{R}^n$.
A ``differentiable manifold'' $M_D$ is a topological manifold equipped with an atlas $\{(U_i,\alpha_i)\}$,
whose transition functions between the images $\alpha_i$, $\alpha_j$ of two overlapping open sets $U_i$, $U_j$ are $C^{\infty}$-differentiable in
$\mathbb{R}^n$.
Here, $\{U_i\}$ is an open covering of $M_D$ and $\alpha_i$ is a homeomorphism, called coordinate,
from the open set $U_i$ onto an open set $U_i'$ of $\mathbb{R}^n$.
Finally, a ``smooth Lorentzian (pseudo-Riemannian) manifold''
$M_L$ is a differentiable
manifold equipped with a smooth metric $g_{\mu\nu}$  of signature $(-+++)$.
For later reference, a summary of the manifolds considered
is given in Table~\ref{tab:Manifolds}, where some of the entries
will be explained later on.

\begin{table}[h]  
\renewcommand{\tabcolsep}{0.25pc}   
\renewcommand{\arraystretch}{1.1}   
\centering
\begin{tabular}{l|c|c|c}   
\text{Manifold}&
\text{Metric}&
\text{Solution of vacuum Einstein eqs.}  
&
\text{Comments}\\
\hhline{=|=|=|=}
$M$&
\text{Lorentzian}&
\text{Yes}&
\text{Minkowski spacetime}\\
\hline
$\widehat{M}_{b}$\;\;&
\text{Ill-defined} &
\text{No}&
\text{Obtained by surgery on $M$}\\
\hline
$\widetilde{\mathcal{M}}_{b}$&
\text{Nonsmooth} &
\text{No}&
\text{Singular Ricci scalar}\\
\hline
$\mathcal{M}_{b}$&
\text{Non-Lorentzian} &
\text{Yes} &
\text{Nonstandard elementary flatness}\\
\hline
\end{tabular}
\caption{Manifolds considered in this article, with defect length scale $b>0$.}
\label{tab:Manifolds}
\end{table}

\subsection{Defect in Minkowski spacetime}
\label{subsec:defect_in_minkowski_space}

\begin{figure}[t] 
\centering                   
\includegraphics[scale=0.7]{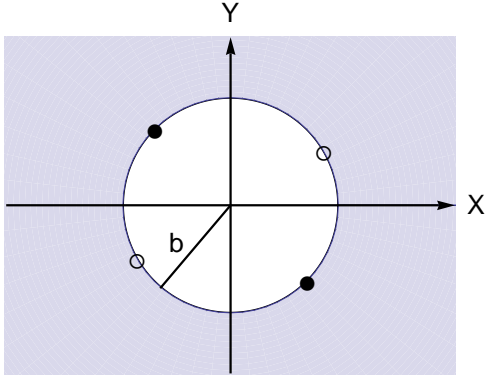}
\caption{Equatorial section ($Z=0$) of the submanifold $\widehat{M}_{b}^{(3)}$. The interior of the sphere of radius $b$ centered at the origin is removed from $\mathbb{R}^3$
and antipodal points on the boundary are identified.}\label{fig:1}
\end{figure}

Consider Minkowski spacetime
with the standard metric for Cartesian coordinates:
\begin{subequations}
\begin{eqnarray}
M   &=& \mathbb{R}\times\mathbb{R}^3\,,\\[2mm]
\eta_{\mu\nu}     &=& \text{diag}(-1,1,1,1)\,,\\[2mm]
x^{\mu}&=&(x^0,x^i) = (x^0,\vec{x})=(t,X,Y,Z)\,,
\end{eqnarray}
\end{subequations}
setting $c=1$.
The spacetime defect of Ref.~\cite{BernadotteKlinkhamer2006}
is obtained by removing a ball of radius $b$ from the spatial hypersurface $\mathbb{R}^3$
and identifying antipodal points on the boundary.
After this surgery, Minkowski spacetime is replaced by the manifold
\begin{subequations}
\begin{eqnarray}
\widehat{M}_{b}&=&\mathbb{R} \times \widehat{M}_{b}^{(3)}\,,
\end{eqnarray}
where the 3-dimensional manifold is given by
\begin{eqnarray}
\widehat{M}_{b}^{(3)}  &=&
\left\{\vec{x}\in\mathbb{R}^3\left|\;|\vec{x}|^2\geq b^2
\wedge
\big(\vec{x} \;\widehat{=}
-\vec{x} \;\;\text{for}\;\;|\vec{x}|^2=b^2\big)\right.\right\}\,,
\end{eqnarray}
\end{subequations}
with the origin of the coordinates $x^i$ chosen to coincide with the center of the defect and the symbol `$\widehat{=}$' standing for pointwise identification.
The structure of this defect is illustrated in Fig.~\ref{fig:1}.

In spherical coordinates
\begin{equation}\label{eq:spher_coord}
(X,Y,Z)=(r\sin\theta\cos\varphi,r\sin\theta\sin\varphi,r\cos\theta)\,,
\end{equation}
the defect is described by the standard Minkowski metric
\begin{subequations}\label{eq:Min_metric_def_spher-AND-def_bound}
\begin{equation}\label{eq:Min_metric_def_spher}
ds^2\,\Big|^{(\widehat{M}_{b})}=
-dt^2+dr^2+r^2\left(d\theta^2+\sin^2\theta\,d\varphi^2\right)\,,
\end{equation}
to which we must append the boundary conditions
\begin{equation}\label{eq:Min_metric_def_bound}
r\geq b>0\,,
\hspace{1cm}
(t,b,\theta,\varphi)\;\widehat{=}\;(t,b,\pi-\theta,\varphi+\pi)\,,
\end{equation}
\end{subequations}
where  the $\varphi$ coordinate is calculated modulo $2\pi$.
The coordinates \eqref{eq:spher_coord} are, however, inappropriate
for the manifold $\widehat{M}_{b}$:
certain points of the manifold  (i.e., those on the defect ``boundary''
at $|\vec{x}|=b$)
have two sets of coordinates and the corresponding metric is
ill-defined for $\widehat{M}_{b}$
(nonzero short distance between identified points).

In order to obtain the topology of this
manifold~\cite{BernadotteKlinkhamer2006,Schwarz2010},
we observe that by using the diffeomorphism ($r\geq b>0$)
\begin{equation}
r\rightarrow\rho=\frac{b}{r}\,,
\end{equation}
it is possible to map the entire manifold $\widehat{M}_{b}^{(3)}$ into the closed unit ball with antipodal
points on the boundary identified
(minus the origin corresponding to spatial infinity).
Since the closed ball with antipodal points on the boundary identified has the
topology of the $3$-dimensional real projective space $\mathbb{R}P^3$, we conclude that the topology of the defect manifold is
\begin{equation}\label{eq:topology}
\widehat{M}_{b}\simeq\mathbb{R}\times\left(\mathbb{R}P^3-\{\text{point}\}\right)\,,
\end{equation}
where `$\simeq$' denotes a homeomorphism.

\subsection{Nonsmooth defect manifold}
\label{subsec:nonsmooth_defect}

In Ref.~\cite{Schwarz2010}, a system of coordinates $\{\widetilde{y},z,x\}$
has been proposed, which is suitable to describe 
the spacetime defect introduced in Sec.~\ref{subsec:defect_in_minkowski_space}.  
We refer to the defect manifold in this system of coordinates as $\widetilde{\mathcal{M}}_{b}$.
In order to completely cover spacetime in this coordinate system, we need to introduce
three charts $U_i$, each one surrounding one of the Cartesian axes $x^i$ but not
intersecting the others. We attach a subscript `$i$' to the new coordinates to indicate to which particular chart it refers:
$\{\widetilde{y}_i,z_i,x_i\}$ is the system of coordinates associated to the chart $U_i$ which surrounds the Cartesian axis $x^i$.

This new system of coordinates is related to the 
standard spherical coordinates \eqref{eq:spher_coord}  
by the following transformations in the first two charts:                                     
\begin{itemize}
\item Chart $U_1$ surrounding $x^1$:
\begin{equation}\left(\begin{split}\label{eq:transformation_U1}
&\widetilde{y}_1=r-b\\
&z_1=\theta\\
&x_1=\varphi\\
\end{split}\right)\hspace{0.5cm}\text{for\;\;\;}|\varphi|<\frac{\pi}{2}\,,\hspace{1cm}
\left(\begin{split}
&\widetilde{y}_1=b-r\\
&z_1=\pi-\theta\\
&x_1=\varphi-\pi\\
\end{split}\right)\hspace{0.5cm}\text{for\;\;\;}|\varphi|>\frac{\pi}{2}\,.\end{equation}
\begin{equation}\left(\begin{split}\label{eq:inverse_transformation_U1}
&r=b+\widetilde{y}_1\\
&\theta=z_1\\
&\varphi=x_1\\
\end{split}\right)\hspace{0.5cm}\text{for\;\;\;}\widetilde{y}_1>0\,,\hspace{1cm}
\left(\begin{split}
&r=b-\widetilde{y}_1\\
&\theta=\pi-z_1\\
&\varphi=x_1+\pi\\
\end{split}\right)\hspace{0.5cm}\text{for\;\;\;}\widetilde{y}_1<0\,.\end{equation}
\item Chart $U_2$ surrounding $x^2$:
\begin{equation}\left(\begin{split}\label{eq:transformation_U2}
&\widetilde{y}_2=r-b\\
&z_2=\theta\\
&x_2=\varphi-\frac{\pi}{2}\\
\end{split}\right)\hspace{0.5cm}\text{for\;\;\;}0<\varphi<\pi\,,\hspace{1cm}
\left(\begin{split}
&\widetilde{y}_2=b-r\\
&z_2=\pi-\theta\\
&x_2=\varphi-\frac{3\pi}{2}\\
\end{split}\right)\hspace{0.5cm}\text{for\;\;\;}\pi<\varphi<2\pi\,.\end{equation}
\begin{equation}\left(\begin{split}\label{eq:inverse_transformation_U2}
&r=b+\widetilde{y}_2\\
&\theta=z_2\\
&\varphi=x_2+\frac{\pi}{2}\\
\end{split}\right)\hspace{0.5cm}\text{for\;\;\;}\widetilde{y}_2>0\,,\hspace{1cm}
\left(\begin{split}
&r=b-\widetilde{y}_2\\
&\theta=\pi-z_2\\
&\varphi=x_2+\frac{3\pi}{2}\\
\end{split}\right)\hspace{0.5cm}\text{for\;\;\;}\widetilde{y}_2<0\,.\end{equation}
\end{itemize}  
Since the standard spherical coordinates are ill defined on the $x^3$ axis,  
it turns out to be useful for the chart $U_3$
to introduce an alternative set of spherical coordinates
\begin{equation}
(X,Y,Z)=(r\sin\hat\theta\sin\hat\varphi,r\cos\hat\theta,r\sin\hat\theta\cos\hat\varphi)\,.
\end{equation}
With this definition, the transformation rules between $\{\widetilde{y}_3,z_3,x_3\}$ and $\{r,\hat\theta,\hat\varphi\}$ are given by
\begin{itemize} 
\item Chart $U_3$ surrounding $x^3$:
\begin{equation}\left(\begin{split}\label{eq:transformation_U3}
&\widetilde{y}_3=r-b\\
&z_3=\hat\theta\\
&x_3=\hat\varphi\\
\end{split}\right)\hspace{0.5cm}\text{for\;\;\;}|\hat\varphi|<\frac{\pi}{2}\,,\hspace{1cm}
\left(\begin{split}
&\widetilde{y}_3=b-r\\
&z_3=\pi-\hat\theta\\
&x_3=\hat\varphi-\pi\\
\end{split}\right)\hspace{0.5cm}\text{for\;\;\;}|\hat\varphi|>\frac{\pi}{2}\,.\end{equation}
\begin{equation}\left(\begin{split}\label{eq:inverse_transformation_U3}
&r=b+\widetilde{y}_3\\
&\hat\theta=z_3\\
&\hat\varphi=x_3\\
\end{split}\right)\hspace{0.5cm}\text{for\;\;\;}\widetilde{y}_3>0\,,\hspace{1cm}
\left(\begin{split}
&r=b-\widetilde{y}_3\\
&\hat\theta=\pi-z_3\\
&\hat\varphi=x_3+\pi\\
\end{split}\right)\hspace{0.5cm}\text{for\;\;\;}\widetilde{y}_3<0\,.\end{equation}
\end{itemize}

\begin{figure}[t]  
\centering
\subfloat[$|\varphi|<\pi/2$]{\includegraphics[scale=0.65]{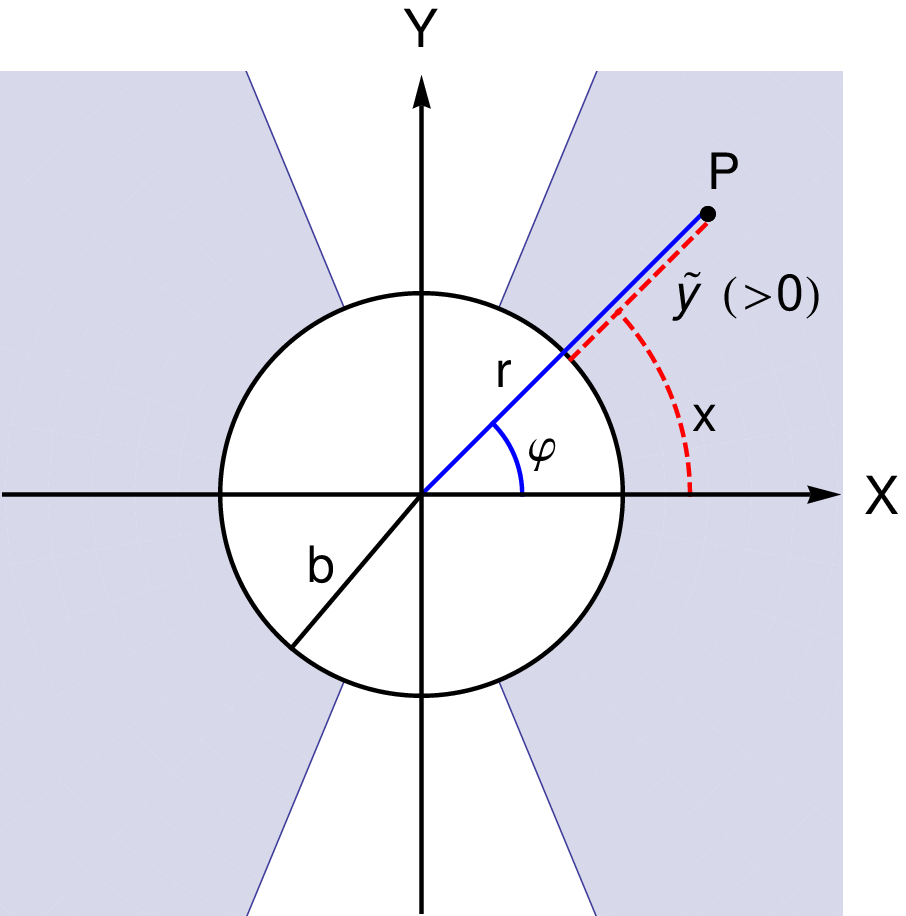}}
\hspace{2cm}
\subfloat[$|\varphi|>\pi/2$]{\includegraphics[scale=0.65]{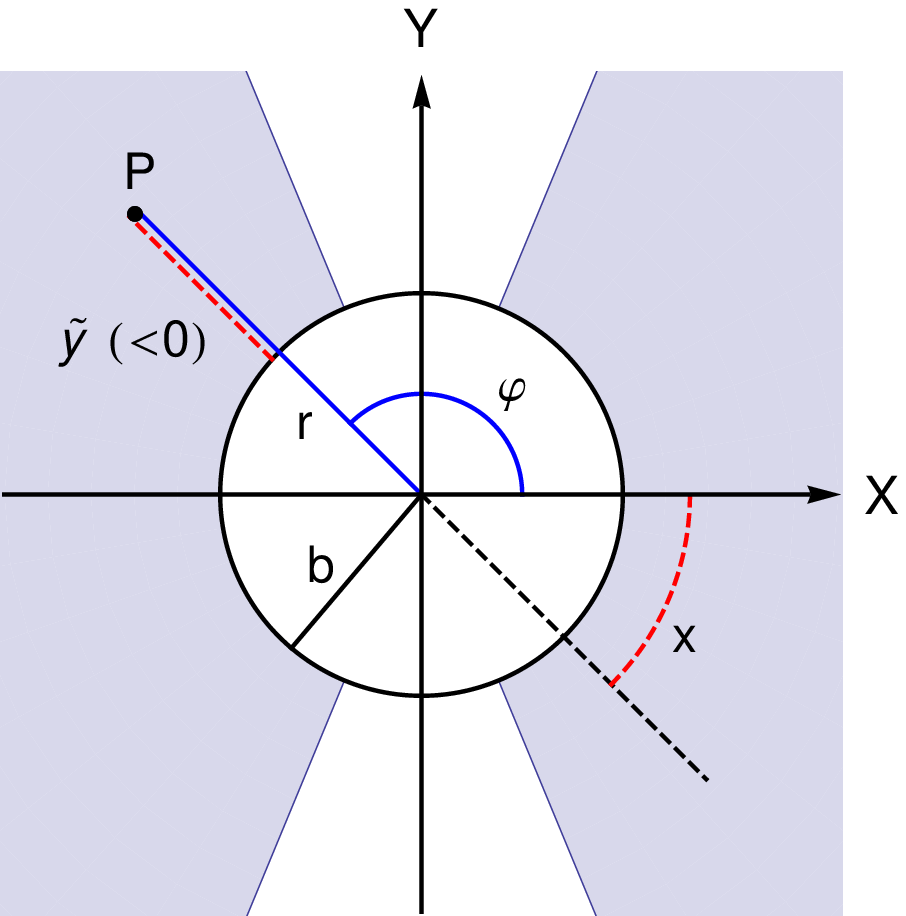}}
\caption{Equatorial section of the
nonsmooth  defect submanifold $\widetilde{\mathcal{M}}_{b}^{(3)}$.
Blue solid lines  indicate the standard spherical coordinates
$(r,\,\varphi)$ of point $P$, with the further spherical coordinate
$\theta=\pi/2$ for the equatorial section. Red dashed lines indicate
the new coordinates $(\widetilde{y},\,x)$ of $P$, with the further
new coordinate $z=\pi/2$ for the equatorial section.
The shaded area represents the chart $U_1$,
where the new coordinate system $\{\widetilde{y},\, z,\, x\}$ is valid.
The smooth  defect submanifold $\mathcal{M}_{b}^{(3)}$ is described
by a similar coordinate system $\{y,\, z,\, x\}$.
}\label{fig:2}
\end{figure}

The standard spherical coordinates \eqref{eq:spher_coord} range over  
\begin{equation}\label{eq:spher_coord_range}
r\in\left[0,\, +\infty\right),\hspace{0.5cm}
\theta\in\left[0,\, \pi\right],\hspace{0.5cm}
\varphi\in\left[0,\, 2\pi\right)\,,
\end{equation}
while the new set of coordinates $\{\widetilde{y},z,x\}$ has ranges
\begin{equation}\label{eq:non_smooth_coord_ranges}
\widetilde{y}\in\left(-\infty,\, +\infty\right),\hspace{0.5cm}
z\in\left(0,\, \pi\right),
\hspace{0.5cm}x\in\left(-\pi/2,\, \pi/2\right)\,,
\end{equation}
for all three charts.
We observe that the angular coordinates $z$ and $x$ cover only half of the solid angle covered by $\theta$ and $\varphi$.
The ``radial'' coordinate $\widetilde{y}$ takes value on the whole real line $\mathbb{R}$, while $r$ only covers the positive real numbers $\mathbb{R}^+$.
See Fig. \ref{fig:2} for a comparison of the two systems of coordinates.
With this choice of coordinates, it is possible to assign a proper atlas to the defect~\cite{Schwarz2010} and $\widetilde{\mathcal{M}}_{b}$ can be regarded
as a differentiable manifold.

The metric of the manifold $\widetilde{\mathcal{M}}_{b}$ can be obtained applying the change of coordinates introduced above
to the defect metric in Minkowski spacetime
\eqref{eq:Min_metric_def_spher}. We then arrive at~\cite{Schwarz2010}
\begin{equation}\label{eq:nonsmooth-def-metric}
ds^2\,\Big|^{(\widetilde{\mathcal{M}}_{b})}
=-dt^2+d\widetilde{y}^2+(b+|\widetilde{y}|)^2\left(dz^2+\sin^2z\,dx^2\right)\,,
\end{equation}
where the defect length scale has been assumed positive, $b>0$,
and the subscript $i$ labeling the three
charts has been dropped, since the metric is the same on each of them.
As said before, one important feature of these coordinates is that we do not need to implement additional boundary conditions to describe the structure of the defect as was needed with spherical coordinates. Observe, however, that
the presence of the absolute value $|\widetilde{y}|$  
makes the metric \eqref{eq:nonsmooth-def-metric} nondifferentiable  
at the defect surface $\widetilde{y}=0$ and  
$\widetilde{\mathcal{M}}_{b}$ cannot be considered a smooth Lorentzian manifold.
This nondifferentiability of the metric
affects~\cite{Schwarz2010} the Ricci scalar $R$ and
the Kretschmann scalar $K$, which are given
by~\cite{Endnote1}%
\begin{equation}\label{eq:non_smooth_curvature_scalars1}
R\equiv g^{\mu\nu}R_{\mu\nu}=-8\,\frac{\delta(\widetilde{y})}{b+|\widetilde{y}|}\,,
\hspace{1cm}K\equiv R_{\mu\nu\rho\sigma}R^{\mu\nu\rho\sigma}=\frac{1}{2}\,R^2\,,
\end{equation}
where $R_{\mu\nu}$ is the Ricci curvature tensor and $R_{\mu\nu\rho\sigma}$ is the Riemann curvature tensor. From these quantities we see that the metric
\eqref{eq:nonsmooth-def-metric} is flat everywhere apart from the defect surface, where the curvature invariants have singularities.

The changes of coordinates given by Eqs. \eqref{eq:transformation_U1}--\eqref{eq:inverse_transformation_U3} define a homeomorphism but not a diffeomorphism between the two defect manifolds
\begin{equation}
\widetilde{\mathcal{M}}_{b}\simeq \widehat{M}_{b}\,,\hspace{1cm}\widetilde{\mathcal{M}}_{b}\not\approx \widehat{M}_{b}\,,
\end{equation}
where `$\approx$' denotes a diffeomorphism.
These transformations (and their inverses) are continuous through the defect, but
the transformation rule for the radial coordinate,
\begin{equation}\label{eq:non_smooth_rad_change_of_coord}
r(\widetilde{y})=b+|\widetilde{y}|\,,
\end{equation}
is nondifferentiable at $\widetilde{y}=0$.   
On the other hand, the two manifolds $\widehat{M}_{b}$ 
and $\widetilde{\mathcal{M}}_{b}$ are locally diffeomorphic 
in the two separate regions $\widetilde{y}>0$ and $\widetilde{y}<0$.  

A last remark concerns the case $b=0$, for which the manifold $\widetilde{\mathcal{M}}_{0}$ describes standard Minkowski
spacetime~\cite{Endnote2}
in coordinates $\{t,\widetilde{y},z,x\}$.
In this case, the metric \eqref{eq:nonsmooth-def-metric} reduces to
\begin{equation}\label{eq:nonsmooth-metric-b0}
ds^2=-dt^2+d\widetilde{y}^2+\widetilde{y}^2\left(dz^2+\sin^2z\,dx^2\right)\,,
\end{equation}
which is smooth everywhere.
In fact, the manifold is now diffeomorphic to Minkowski
spacetime~\cite{Endnote3}
\begin{equation}
\widetilde{\mathcal{M}}_{0}\approx M\,.
\end{equation}

\subsection{Smooth defect manifold}
\label{subsec:smooth_defect}

We next introduce the third type of spacetime defect $\mathcal{M}_{b}$.
This defect manifold
has been obtained in Ref.~\cite{KlinkhamerRahmede2013} as a vacuum solution of general relativity.
It can be described in a coordinate system $\{t,y,z,x\}$ analogous to the one introduced in 
Sec.~\ref{subsec:nonsmooth_defect}, since it also has ranges
\begin{equation}\label{eq:smooth_coord_ranges}
t\in\left(-\infty,\, +\infty\right),\hspace{0.5cm}
y\in\left(-\infty,\, +\infty\right),\hspace{0.5cm}
z\in\left(0,\, \pi\right),\hspace{0.5cm}
x\in\left(-\pi/2,\, \pi/2\right)\,,
\end{equation}
but is differently related to the standard spherical coordinates
(see below for details). Figure~\ref{fig:2} also 
gives a sketch of $\mathcal{M}_{b}$, with $\widetilde{y}$ replaced by $y$
in the panels. 

In these coordinates, the defect metric takes the form
\begin{eqnarray}\label{eq:smooth_mass_def_metric}
ds^2&=&
-\left(1-\frac{\ell}{\sqrt{y^2+b^2}}\right)\,dt^2
+\left(1-\frac{\ell}{\sqrt{y^2+b^2}}\right)^{-1}\, \frac{y^2}{y^2+b^2}\,dy^2
\nonumber\\[1mm]
&&
+(y^2+b^2)\,(dz^2+\sin^2z\,dx^2)\,,
\end{eqnarray}
where $b>0$ gives the defect length scale
and the additional length parameter $\ell<b$ is related to
the defect mass by $m=\ell/(2\,G)$, recalling that we have set $c=1$.
In this article, we only consider the case of a massless defect ($\ell=0$), for which the defect metric
simplifies to
\begin{equation}\label{eq:smooth_def_metric}
ds^2\,\Big|^{(\mathcal{M}_{b})}
=-dt^2+\frac{y^2}{(y^2+b^2)}\,dy^2+(y^2+b^2)\,(dz^2+\sin^2z\,dx^2)\,.
\end{equation}
We see that this metric is smooth and well-behaved
everywhere~\cite{Endnote4}.
However, $\mathcal{M}_{b}$  is not a genuine Lorentzian manifold,
because it has nonstandard elementary flatness at certain points (see below).

The crucial difference with respect to the defect $\widetilde{\mathcal{M}}_{b}$
is the relation between the new set of coordinates
\eqref{eq:smooth_coord_ranges} and the standard spherical coordinates.
For $\mathcal{M}_b$, the radial coordinate $r$ is related to $y$ by the
equation~\cite{Endnote5}
\begin{equation}\label{eq:smooth_rad_change_of_coord}
r=\sqrt{y^2+b^2}\,,\hspace{1cm}\forall y\,,
\end{equation}
whose inverse is
\begin{equation}\label{eq:smooth_rad_change_of_coord_inverse}
y=\left\{\begin{split}&+\sqrt{r^2-b^2}\,,\hspace{1cm}
&&\text{for\;\;\;}|\varphi|<\pi/2\,,\\
&-\sqrt{r^2-b^2}\,,\hspace{1cm}
&&\text{for\;\;\;}|\varphi|>\pi/2\,.\\
\end{split}\right.
\end{equation}
Observe that this statement can be regarded as an \textit{a posteriori} conclusion. In fact, the metric \eqref{eq:smooth_def_metric} can be obtained in the coordinate system $\{t,y,z,x\}$ without any knowledge of its relation to spherical coordinates. Then, noting that far away from the defect surface this metric must be equivalent to Minkowski metric, Eq. \eqref{eq:smooth_rad_change_of_coord} must hold.

The angular spherical coordinates are related to the coordinates $z$ and $x$ as in 
Sec.~\ref{subsec:nonsmooth_defect}.  
Explicitly, the change of coordinates in the chart $U_1$ surrounding
the Cartesian axis $x^1$ reads
\begin{equation}\left(\begin{split}\label{eq:transformation_U1_smooth}
&y_1=\sqrt{r^2-b^2}\\
&z_1=\theta\\
&x_1=\varphi\\
\end{split}\right)\hspace{0.5cm}\text{for\;\;\;}|\varphi|<\frac{\pi}{2}\,,\hspace{1cm}
\left(\begin{split}
&y_1=-\sqrt{r^2-b^2}\\
&z_1=\pi-\theta\\
&x_1=\varphi-\pi\\
\end{split}\right)\hspace{0.5cm}\text{for\;\;\;}|\varphi|>\frac{\pi}{2}\,.\end{equation}
\begin{equation}\left(\begin{split}\label{eq:inverse_transformation_U1_smooth}
&r=\sqrt{y_1^2+b^2}\\
&\theta=z_1\\
&\varphi=x_1\\
\end{split}\right)\hspace{0.5cm}\text{for\;\;\;}y_1>0\,,\hspace{1cm}
\left(\begin{split}
&r=\sqrt{y_1^2+b^2}\\
&\theta=\pi-z_1\\
&\varphi=x_1+\pi\\
\end{split}\right)\hspace{0.5cm}\text{for\;\;\;}y_1<0\,,\end{equation}
and similarly for the two other charts, $U_2$ and $U_3$
[obtained by making the obvious changes in
\eqref{eq:transformation_U2}--\eqref{eq:inverse_transformation_U3}].

As for the previous type of defect, this system of coordinates automatically implements the antipodal identification on the boundary of the defect. The defect manifold $\mathcal{M}_{b}$ is completely determined by the metric \eqref{eq:smooth_def_metric} without need to introduce additional boundary conditions. In this case, the Ricci and Kretschmann scalars turn out to be regular everywhere
\begin{equation}\label{eq:smooth_curvature_scalars}
R\equiv g^{\mu\nu}R_{\mu\nu}=0\,,\hspace{1cm}
K\equiv R_{\mu\nu\rho\sigma}R^{\mu\nu\rho\sigma}=\frac{12\,\ell^2}{(y^2+b^2)^3}\,,
\end{equation}
where we have temporarily considered the general metric \eqref{eq:smooth_mass_def_metric} with nonvanishing parameter $\ell$.

Observe that the smoothness of the manifold $\mathcal{M}_{b}$
comes at the price of relaxing the
standard elementary-flatness condition~\cite{Klinkhamer2013,Klinkhamer2013-review}.
It is possible to transform the metric \eqref{eq:smooth_def_metric}
in a neighborhood of the defect boundary $y=0$ to the standard Minkowski metric.
However, the coordinate transformation
is a $C^1$ function and not a $C^\infty$-diffeomorphism, as required by the
standard elementary flatness condition.
Remark also that the matrix of the metric $g_{\mu\nu}(t,\,x,\,y,\,z)$
from \eqref{eq:smooth_mass_def_metric} has a vanishing determinant at $y=0$,
i.e., the metric is degenerate there, contrary to the assumptions of
standard
general
relativity~\cite{Einstein1916,HawkingEllis1973}.
For this reason, the metric \eqref{eq:smooth_mass_def_metric} is
non-Lorentzian and has been called a ``regularization'' of the
standard Schwarzschild metric~\cite{Klinkhamer2013-review}.
Just to be clear, the metric \eqref{eq:smooth_mass_def_metric} is a solution
in general relativity but \emph{not}  in \emph{standard} general relativity
(this correct a statement in the last paragraph of Sec.~1 of
Ref~\cite{Klinkhamer2013-review}).

Specifically, the change of coordinates between the systems $\{y,z,t\}$
and $\{\widetilde{y},z,t\}$ is given by
\begin{equation}y=\left\{\begin{split}\label{eq:transformation_y_y}
&+\sqrt{\widetilde{y}^2+2b|\widetilde{y}|}\,,
\hspace{1cm}\text{for\;\;\;}\widetilde{y}>0\,,\\
&-\sqrt{\widetilde{y}^2+2b|\widetilde{y}|}\,,\hspace{1cm}\text{for\;\;\;}\widetilde{y}<0\,,
\end{split}\right.\hspace{1cm}
\widetilde{y}=\left\{\begin{split}
&\sqrt{y^2+b^2}-b\,,\hspace{1cm}\text{for\;\;\;}y>0\,,\\
&b-\sqrt{y^2+b^2}\,,\hspace{1cm}\text{for\;\;\;}y<0\,.
\end{split}\right.
\end{equation}
The defect manifold $\mathcal{M}_{b}$ is homeomorphic but not diffeomorphic to the previous types of defects
\begin{subequations}
\begin{eqnarray}
\mathcal{M}_{b}&\simeq& \widetilde{\mathcal{M}}_{b}\simeq\widehat{M}_{b}\,,\label{eq:homeomorphism_y_y1_r}\\[2mm]
\mathcal{M}_{b}&\not\approx& \widehat{M}_{b}\,,\label{eq:non_diff_y_r}\\[2mm]
\mathcal{M}_{b} &\not\approx& \widetilde{\mathcal{M}}_{b}\,.\label{eq:non_diff_y_y1}
\end{eqnarray}
\end{subequations}
The homeomorphism relation \eqref{eq:homeomorphism_y_y1_r} is simply proved observing that both transformations given by
Eqs. \eqref{eq:transformation_U1_smooth} and \eqref{eq:inverse_transformation_U1_smooth} for $\mathcal{M}_{b}\leftrightarrow \widehat{M}_{b}$
and by Eq. \eqref{eq:transformation_y_y} for $\mathcal{M}_{b}\leftrightarrow \widetilde{\mathcal{M}}_{b}$ are
continuous and have continuous inverses. The nondiffeomorphism relations \eqref{eq:non_diff_y_r} and \eqref{eq:non_diff_y_y1}
follow from the fact that these transformations (or their inverses) are not $C^{\infty}$-differentiable at the defect surface $y=0$.
Again, in the two separate regions $y>0$ and $y<0$, the three manifolds are locally diffeomorphic.

Considering the case $b=0$, we observe that the metric \eqref{eq:smooth_def_metric} reduces to Eq. \eqref{eq:nonsmooth-metric-b0}, while the change of
coordinates \eqref{eq:transformation_y_y} reduces to $y=\widetilde{y}$. Then, we can state that
\begin{equation}
\mathcal{M}_{0}=\widetilde{\mathcal{M}}_{0}\approx M\,.
\end{equation}
The main focus of this article will, however, be on the case $b\ne 0$.

\section{Scalar field solutions}
\label{sec:scalar_solutions}

\subsection{General solution}
\label{subsec:General-solution}

Since the defect manifolds of Sec.~\ref{sec:framework} 
have distinct differential structures, we expect the physics (governed by differential
equations) also to be different. To show this explicitly, we examine the case of a massless scalar field.

The massless Klein--Gordon equation for a real scalar field $\mathbf{\Phi}$ in a general metric $g_{\mu\nu}$ is given by~\cite{BirrellDavies1984}
\begin{equation}\label{eq:klein_gordon_general}
\Box\mathbf{\Phi}
\equiv \nabla_{\mu}\nabla^{\mu}\mathbf{\Phi}
=g^{-1/2}\,\partial_{\mu}
\left(g^{1/2}\,g^{\mu\nu}\partial_{\nu}\mathbf{\Phi}\right)=0\,,
\end{equation}
where $g\equiv -\det(g_{\mu\nu})$. For our three defect manifolds, this yields
\begin{subequations}\label{eq:klein_gordon}
\begin{align}
&\widehat{M}_{b}:&&-\partial^2_t\mathbf{\Phi}
+\partial^2_r\mathbf{\Phi}+\frac{2}{r}\partial_r\mathbf{\Phi}+
\frac{\partial^2_\theta\mathbf{\Phi}}{r^2}+\frac{\cot \theta\,\partial_\theta\mathbf{\Phi}}{r^2}
+\frac{\partial^2_\varphi\mathbf{\Phi}}{r^2\sin^2\theta}=0\,,
\label{eq:Min_klein_gordon}\\[2mm] &\widetilde{\mathcal{M}}_{b}:&&-\partial^2_t\mathbf{\Phi}
+\partial^2_{\widetilde{y}}\mathbf{\Phi}
+\frac{2\widetilde{y}\partial_{\widetilde{y}}\mathbf{\Phi}}{|\widetilde{y}|
(b+|\widetilde{y}|)}
\nonumber\\&&&
+\frac{\partial^2_z\mathbf{\Phi}}{(b+|\widetilde{y}|)^2}
+\frac{\cot z\,\partial_z\mathbf{\Phi}}{(b+|\widetilde{y}|)^2}+\frac{\partial^2_x\mathbf{\Phi}}{(b+|\widetilde{y}|)^2\sin^2z}=0\,,
\label{eq:non_smooth_def_klein_gordon}\\[2mm]
&\mathcal{M}_{b}:&&-\partial^2_t\mathbf{\Phi}+\frac{y^2+b^2}{y^2}\partial^2_y\mathbf{\Phi}
+\frac{2y^2-b^2}{y^3}\partial_y\mathbf{\Phi}
\nonumber\\&&&
+\frac{\partial^2_z\mathbf{\Phi}}{y^2+b^2}
+\frac{\cot z\,\partial_z\mathbf{\Phi}}{y^2+b^2}+\frac{\partial^2_x\mathbf{\Phi}}{(y^2+b^2)\sin^2z}=0\,,\label{eq:smooth_def_klein_gordon}
\end{align}\end{subequations}
and we recall that Eq. \eqref{eq:Min_klein_gordon}
must be supplemented by boundary conditions corresponding to
\eqref{eq:Min_metric_def_bound}.

In order to find the solutions of the scalar equations \eqref{eq:klein_gordon}, 
we use the standard method of separation of variables,
\begin{equation}
\mathbf{\Phi}(t,r,\theta,\varphi)=
T(t)\, R(r)\, \Theta(\theta)\, \varPhi(\varphi)\,,
\end{equation}
where $r$ must be replaced by $\widetilde{y}$ or $y$ for
$\widetilde{\mathcal{M}}_{b}$ or $\mathcal{M}_{b}$
and similarly $\theta$ and $\varphi$ must be replaced by $z$ and $x$.
We obtain that the time and angular equations are equal for the three cases:
\begin{subequations}\begin{align}
&\partial_t^2T+k^2T=0\,,\label{eq:time_eq}\\[2mm]
&\partial_\theta^2\Theta+\cot\theta\,\partial_\theta\Theta+\left(l(l+1)-\frac{m^2}{\sin^2\theta}\right)\Theta=0\,,\label{eq:theta_eq}\\
&\partial_\varphi^2\varPhi+m^2\varPhi=0\,.\label{eq:phi_eq}
\end{align}\end{subequations}
Hence, the temporal solution turns out to be  
\begin{subequations}\begin{align}\label{eq:time_ang_sol}
&T(t)\propto e^{-i\,\omega_k\,t}\,,&&\omega_k^2=k^2\,,&&
t\in(-\infty,\, +\infty)\,,
\end{align}
for all three defect manifolds considered
($\widehat{M}_{b}$, $\widetilde{\mathcal{M}}_{b}$,
and $\mathcal{M}_{b}$). The angular solutions are
\begin{align}
&\Theta(\theta)\varPhi(\varphi)\propto\left\{\begin{aligned}&Y_{l}^{m}(\theta,\varphi)\,,\\&Y_{l}^{m}(z,x)\,,\end{aligned}\right.
&&\begin{aligned}&\theta\in[0,\, \pi]\,,\varphi\in[0,\, 2\pi)\,,\\
&z\in(0,\, \pi)\,,\,x\in\left(-\pi/2,\, \pi/2\right)\,,
\end{aligned}
&&\begin{aligned}&\text{for\;\;\;}\widehat{M}_{b}\,,\\&\text{for\;\;\;}\widetilde{\mathcal{M}}_{b}\,,\mathcal{M}_{b}\,,\end{aligned}\label{eq:ang_sol}
\end{align}\end{subequations}
where $Y_{l}^{m}(\theta,\varphi)$ are the standard spherical harmonics.

The general solution of Eq. \eqref{eq:Min_klein_gordon} can, then, be written as
\begin{subequations}\label{eq:gen_sol}
\begin{eqnarray}
\mathbf{\Phi}^{(\widehat{M}_{b})}(t,r,\theta,\varphi)&=&
\int dk\sum_{l,m}A_{klm}\mathbf{\Phi}^{(\widehat{M}_{b})}_{klm}(r,\theta,\varphi)e^{-i\omega_k t}\,,\\[2mm]
\mathbf{\Phi}^{(\widehat{M}_{b})}_{klm}(r,\theta,\varphi)&=&
R_{kl}(r)\, Y_{l}^{m}(\theta,\phi)\,,
\end{eqnarray}
\end{subequations}
and similarly for the solutions of Eqs. \eqref{eq:non_smooth_def_klein_gordon} and \eqref{eq:smooth_def_klein_gordon}.
The separation constant $k$ from \eqref{eq:gen_sol} ranges over $\mathbb{R}$.
Its modulus is identified with the wavenumber
$|k|=2\pi/\lambda$, where $\lambda$ is the wavelength of the scalar mode,
and its sign distinguishes between incoming and outgoing modes.
The constants $l$ and $m$ are integers,
$l\geq0$ and $|m|\leq l$. It needs to be emphasized
that these separation constants
($k$, $l$, and $m$) do not change under parity transformations.

The radial solutions $R_{kl}(r)$ are obtained from the radial equations
\begin{subequations}\label{eq:rad_eqs}
\begin{align}
&\widehat{M}_{b}:&&\partial^2_rR_{kl}(r)+\frac{2}{r}\;\partial_rR_{kl}(r)+\left(k^2-\frac{l(l+1)}{r^2}\right)R_{kl}(r)=0\,,
\label{eq:rad_eq_Min}\\[2mm]
&\widetilde{\mathcal{M}}_{b}:&&\partial^2_{\widetilde{y}}R_{kl}(\widetilde{y})
+\frac{2\widetilde{y}}{|\widetilde{y}|(b+|\widetilde{y}|)}\;
\partial_{\widetilde{y}}R_{kl}(\widetilde{y})
+\left(k^2-\frac{l(l+1)}{(b+|\widetilde{y}|)^2}\right)R_{kl}(\widetilde{y})=0\,,
\label{eq:rad_eq_NS}\\[2mm]
&\mathcal{M}_{b}:&&\frac{y^2+b^2}{y^2}\;\partial^2_yR_{kl}(y)
+\frac{2y^2-b^2}{y^3}\;\partial_yR_{kl}(y)
+\left(k^2-\frac{l(l+1)}{y^2+b^2}\right)R_{kl}(y)=0\,.
\label{eq:rad_eq_S}
\end{align}\end{subequations}

Before proceeding to the study of these equations,
we want to state clearly what a solution (or proper solution)
of an ordinary differential equation is.
Following Ref. \cite{TenembaumPollard1963}, a function $f(x)$ can be regarded as a solution of an ordinary differential equation on a domain $I$ if it solves
the equation for every $x\in I$. This means that $f(x)$ must be defined everywhere in the domain $I$ and that, in particular, it cannot be discontinuous,
since, at the discontinuity, a unique value of the function is not defined. The same must be true for the derivatives of $f(x)$, at least up to the order
of the differential equation. Hence, a solution of an ordinary differential equation of order $n$ on a domain $I$ must be, at least, $C^n$-differentiable on $I$.

Observe, then, that proper global solutions of the radial equations \eqref{eq:rad_eqs}
can only be found for the smooth defect manifold $\mathcal{M}_{b}$. In fact, only $\mathcal{M}_{b}$ is a smooth manifold whose differential structure and metric are well defined everywhere. For the other two cases ($\widehat{M}_{b}$ and $\widetilde{\mathcal{M}}_{b}$), we must rely on boundary conditions at the defect surface to find global solutions.

\subsection{Scalar solution over Minkowski spacetime}
\label{subsec:minkowski_space}

We start by considering the simplest case $b=0$. In this case, 
the three manifolds are mutually diffeomorphic
and the radial equations \eqref{eq:rad_eqs} 
become formally equal and coincide with the standard spherical Bessel equation. Then, we obtain the solutions
\begin{subequations}\label{eq:rad_sol_Min}\begin{align}
&R_{kl}(r)=j_l(kr)\,,&&r\in[0,\, +\infty)\,,
&&\text{for\;\;\;}M\,,
\label{eq:rad_sol_Min_r}\\[2mm]
&R_{kl}(\widetilde{y})=R_{kl}(y)=j_l(ky)\,,&&
y\in(-\infty,\, +\infty)\,,
&&\text{for\;\;\;}\widetilde{\mathcal{M}}_{0} =\mathcal{M}_{0}   \,,
\label{eq:rad_sol_Min_y}
\end{align}\end{subequations}
where $j_l(kr)$ is the spherical Bessel function of the first kind
and we do not consider the spherical Bessel function of the second kind $y_l(kr)$ since
it diverges at the origin.

Taking into account also the angular results \eqref{eq:ang_sol}, the field solutions are
\begin{subequations}\label{eq:scal_sol_0}
\begin{align}
&\mathbf{\Phi}^{(M)}_{klm}(r,\theta,\varphi)\;\;
=j_l(kr)\, Y_{l}^{m}(\theta,\phi)\,,\label{eq:scal_sol_0_r}\\[2mm]
&\mathbf{\Phi}^{(\mathcal{M}_{0})}_{klm}(y,z,x)
=j_l(ky)\, Y_{l}^{m}(z,x)\,,\label{eq:scal_sol_0_y}
\end{align}\end{subequations}
which are, indeed, equivalent, since they transform into each other under the changes of coordinates \eqref{eq:transformation_U1_smooth} and
\eqref{eq:inverse_transformation_U1_smooth}. These coordinate transformations
guarantee that the two manifolds $M$ and $\mathcal{M}_{0}$ are locally diffeomorphic in the two regions $y>0$ and $y<0$. However, they do not give information about the origin $y=0$. It is important, then, to verify that the two solutions
behave consistently across this point.

It is useful, in this regard, to compare the behavior of the
solutions under parity. The parity transformation for the different coordinate systems is given by
\begin{align}
\vec{x}=\left\{\begin{aligned}&(X,Y,Z)\\&(r,\theta,\phi)\\&(y,z,x)\end{aligned}\right.
\overset{P}\longrightarrow
-\vec{x}=\left\{\begin{aligned}&(-X,-Y,-Z)\\
&(r,\pi-\theta,\pi+\phi)\\&(-y,z,x)\end{aligned}\right.\hspace{1cm}
\begin{aligned}&\text{Cartesian};\\
&\text{spherical};\\&\text{real-projective}.
\end{aligned}
\end{align}
Applied to the scalar solutions \eqref{eq:scal_sol_0}, this gives
\begin{subequations}\begin{align}
&\mathbf{\Phi}^{(M)}_{klm}(r,\theta,\varphi)\overset{P}\longrightarrow \mathbf{\Phi}^{(M)}_{klm}(r,\pi-\theta,\varphi+\pi)=
(-1)^{l}\, \mathbf{\Phi}^{(M)}_{klm}(r,\theta,\varphi)\,,\\[2mm]
&\mathbf{\Phi}^{(\mathcal{M}_{0})}_{klm}(y,z,x)\overset{P}\longrightarrow \mathbf{\Phi}^{(\mathcal{M}_{0})}_{klm}(-y,z,x)=
(-1)^{l}\, \mathbf{\Phi}^{(\mathcal{M}_{0})}_{klm}(y,z,x)\,.
\end{align}\end{subequations}
We see that the two scalar solutions have the same parity eigenvalues, which, however, have different origins. Specifically, the parity operator for Minkowski spacetime $M$ acts on the angular variables and the $(-1)^{l}$ factor comes from the behavior of the spherical harmonics $Y_{l}^{m}(\theta,\varphi)$. For $\mathcal{M}_{0}$, the parity operator acts on the ``radial'' coordinate $y$ and the $(-1)^{l}$ factor comes from the behavior of the spherical Bessel function $j_l(y)$.

\subsection{Scalar solution for the defect in Minkowski spacetime}
\label{subsec:Minkowski_space_solutions}

For the case of $b\neq0$ in spherical coordinates,
we need to take into account the defect structure \eqref{eq:Min_metric_def_bound}
and to require
that the field is continuous at the defect surface,
\begin{equation}\label{eq:bound_cond_scalar}
\mathbf{\Phi}^{(\widehat{M}_{b})}(t,\,b,\,\theta,\,\varphi)=
\mathbf{\Phi}^{(\widehat{M}_{b})}(t,\,b,\,\pi-\theta,\,\varphi+\pi)\,.
\end{equation}
Such a condition is not satisfied by the standard Minkowski solution $\mathbf{\Phi}^{(M)}(t,\,r,\,\theta,\,\varphi)$ and we need to introduce
an additional scattered field $\mathbf{\Phi}^{(M_S)}(t,\,r,\,\theta,\,\varphi)$, so that the total field,
\begin{equation}
\mathbf{\Phi}^{(\widehat{M}_{b})}(t,\,r,\,\theta,\,\varphi)=
\mathbf{\Phi}^{(M)}(t,\,r,\,\theta,\,\varphi)+\mathbf{\Phi}^{(M_S)}(t,\,r,\,\theta,\,\varphi)\,,
\end{equation}
behaves correctly.

The scattered field $\mathbf{\Phi}^{(M_S)}$
can still be expressed as in Eq. \eqref{eq:gen_sol},  
where the coefficients $A'_{klm}$ are determined by the
boundary condition \eqref{eq:bound_cond_scalar}.
The radial solution $R^{(M_S)}_{kl}(r)$ can, in principle, be any of the spherical Bessel functions.
Imposing the Sommerfeld radiation condition~\cite{Atkinson1949},
which requires that the solution must behaves as $e^{ikr}/r$ at large $r$,
allows us to identify the radial solution $R^{(M_S)}_{kl}(r)$ with the spherical Hankel function $h_l^{(1)}(kr)$. The total solution turns out to be
\begin{equation}\label{eq:scal_sol_Mb_r}
\mathbf{\Phi}^{(\widehat{M}_{b})}_{klm}(r,\theta,\varphi)=
\left[j_l(kr)-\left(\frac{1-(-1)^{l}}{2}\frac{j_l(kb)}{h_l^{(1)}(kb)}\right)
h_l^{(1)}(kr)\right]\, Y_{l}^{m}(\theta,\phi)\,,
\end{equation}
where the explicit dependence on $t$ is dropped.

Using the transformations \eqref{eq:transformation_U1_smooth} we obtain that this radial solution, written in the system of coordinates $\{y,z,x\}$, takes the form
\begin{equation}\label{eq:rad_sol_Mb_y_S}
R_{kl}^{(\widehat{M}_{b})}(y)=\left\{
\begin{split}
&j_l(k\sqrt{y^2+b^2})-\left(\frac{1-(-1)^{l}}{2}\frac{j_l(kb)}{h_l^{(1)}(kb)}\right)h_l^{(1)}(k\sqrt{y^2+b^2})\,,
\hspace{0.5cm}&&\text{for\;\;\;}y>0\,,\\
&j_l(-k\sqrt{y^2+b^2})-\left(\frac{1-(-1)^{l}}{2}\frac{j_l(-kb)}{h_l^{(1)}(-kb)}\right)h_l^{(1)}(-k\sqrt{y^2+b^2})\,,
\hspace{0.5cm}&&\text{for\;\;\;}y<0\,.
\end{split}\right.
\end{equation}
Similarly, from the change of coordinates \eqref{eq:transformation_U1}, we obtain in the system $\{\widetilde{y},z,x\}$
\begin{equation}\label{eq:rad_sol_Mb_y_NS}
R_{kl}^{(\widehat{M}_{b})}(\widetilde{y})=\left\{
\begin{split}
&j_l(k(b+|\widetilde{y}|))-\left(\frac{1-(-1)^{l}}{2}\frac{j_l(kb)}{h_l^{(1)}(kb)}\right)h_l^{(1)}(k(b+|\widetilde{y}|))\,,
\hspace{0.5cm}&&\text{for\;\;\;}\widetilde{y}>0\,,\\
&j_l(-k(b+|\widetilde{y}|))-\left(\frac{1-(-1)^{l}}{2}\frac{j_l(-kb)}{h_l^{(1)}(-kb)}\right)h_l^{(1)}(-k(b+|\widetilde{y}|))\,,
\hspace{0.5cm}&&\text{for\;\;\;}\widetilde{y}<0\,.
\end{split}\right.
\end{equation}

The solution \eqref{eq:scal_sol_Mb_r} for odd values of $l$ manifestly vanishes at $r=b$. The same holds for the transformed  solutions
\eqref{eq:rad_sol_Mb_y_S} and \eqref{eq:rad_sol_Mb_y_NS} at, respectively, $y=0$ and $\widetilde{y}=0$.

\subsection{Scalar solution for the smooth defect}
\label{subsec:smooth_defect_solutions}

We now consider the smooth defect manifold $\mathcal{M}_{b}$ described by the metric \eqref{eq:smooth_def_metric} for $b>0$. In this case, the radial solution
of Eq. \eqref{eq:rad_eq_S} is
\begin{equation}\label{eq:rad_sol_MbD_y}
R_{kl}^{(\mathcal{M}_{b})}(y)=j_l(k\sqrt{y^2+b^2})\,,
\end{equation}
from which follows the scalar solution
\begin{equation}\label{eq:scal_sol_MbD_y}
\mathbf{\Phi}^{(\mathcal{M}_{b})}_{klm}(y,z,x)=j_l(k\sqrt{y^2+b^2})\, Y_{l}^{m}(z,x)\,.
\end{equation}
In principle, a second independent radial solution proportional to the spherical Bessel function of the
second kind is allowed. However, for simplicity, we neglect this second solution by requiring that the total radial function must approach the standard
Minkowski result \eqref{eq:rad_sol_Min_y} as $y\rightarrow +\infty$ . The conclusions are not affected by this restriction.

Several remarks are in order.
First, the solution \eqref{eq:rad_sol_MbD_y}
is a proper global solution of the radial equation \eqref{eq:rad_eq_S} only for $b>0$. Setting $b=0$ in Eq. \eqref{eq:rad_sol_MbD_y} gives
a function which is nondifferentiable at $y=0$ (and cannot be regarded as a proper solution as defined at the end of Sec. \ref{subsec:General-solution}).
This has to be expected since, as we have shown in Sec. \ref{subsec:smooth_defect}, the properties of the manifolds $\mathcal{M}_{b}$ and $\mathcal{M}_{0}$ are
fundamentally different: $\mathcal{M}_{0}$ is diffeomorphic to Minkowski spacetime, whereas $\mathcal{M}_{b}$ is not. Studying the radial Klein--Gordon
equation for $\mathcal{M}_{0}$ we obtain, in fact, the solution \eqref{eq:rad_sol_Min_y}, which is differentiable at $y=0$.

Still,  the smooth manifold $\mathcal{M}_{b}$ for $b>0$
is locally diffeomorphic to Minkowski spacetime in the two separate regions $y>0$ and $y<0$.
Then, we expect the solution \eqref{eq:scal_sol_MbD_y} to be equivalent to the one obtained for $M$ in these two regions.
Applying the change of coordinate \eqref{eq:transformation_U1_smooth} to the solution \eqref{eq:scal_sol_0_r} we obtain
\begin{equation}
\mathbf{\Phi}^{(M)}_{klm}(y,z,x)=\left\{\begin{split}\label{eq:scal_M0_by}
&j_l(+k\sqrt{y^2+b^2})\, Y_{l}^{m}(z,x)\,,\hspace{1cm}&&\text{for\;\;\;}y>0\,,\\
&j_l(-k\sqrt{y^2+b^2})\, Y_{l}^{m}(z,x)\,,\hspace{1cm}&&\text{for\;\;\;}y<0\,.
\end{split}\right.
\end{equation}
This last expression indeed coincides with Eq. \eqref{eq:scal_sol_MbD_y} in the region $y>0$ but not in the region $y<0$.
Note that Eq. \eqref{eq:scal_M0_by} cannot be regarded as a solution of the Klein--Gordon equation
in $\mathcal{M}_b$ because it is discontinuous at $y=0$. This shows explicitly that $\mathcal{M}_{b}$ and $M$ are not globally diffeomorphic.

We must also compare $\mathcal{M}_{b}$ with $\widehat{M}_{b}$. To do so, we investigate if the scalar solution obtained for $\widehat{M}_{b}$,
Eq. \eqref{eq:scal_sol_Mb_r}, can be a proper solution for the smooth manifold $\mathcal{M}_{b}$. We have already transformed this equation into
the system of coordinates $\{y,z,x\}$ and the resulting radial function $R_{kl}^{(\widehat{M}_{b})}(y)$ is given by Eq. \eqref{eq:rad_sol_Mb_y_S}.
We observe that this expression is continuous at $y=0$ and, inserted into the radial equation \eqref{eq:rad_eq_S}, turns out to be a solution in both
regions $y>0$ and $y<0$. However, its second derivative is discontinuous at $y=0$
and, consequently, it cannot be considered a proper solution at the defect surface.
Again, this shows explicitly that $\mathcal{M}_{b}$ and $\widehat{M}_{b}$ are not globally diffeomorphic.

Just as for the $b=0$ case discussed in Sec.~\ref{subsec:minkowski_space},
it is useful to study the behavior of the solutions under parity. We observe that
\begin{subequations}\begin{align}
&\mathbf{\Phi}^{(\widehat{M}_{b})}_{klm}(r,\theta,\varphi)\overset{P}\longrightarrow     \mathbf{\Phi}^{(\widehat{M}_{b})}_{klm}(r,\pi-\theta,\varphi+\pi)
=(-1)^{l}\mathbf{\Phi}^{(\widehat{M}_{b})}_{klm}(r,\theta,\varphi)\,,\\[2mm]
&\mathbf{\Phi}^{(\mathcal{M}_{b})}_{klm}(y,z,x)\overset{P}\longrightarrow
\mathbf{\Phi}^{(\mathcal{M}_{b})}_{klm}(-y,z,x)
=(+1)\mathbf{\Phi}^{(\mathcal{M}_{b})}_{klm}(y,z,x)\,.
\end{align}\end{subequations}
Hence, the solution for the defect manifold $\mathcal{M}_{b}$ 
behaves differently compared to the solution for the defect in Minkowski spacetime.  
This shows, once more, that, the scalar solutions obtained for the 
two manifolds $\widehat{M}_{b}$ and $\mathcal{M}_{b}$, 
with $b\neq0$, are inequivalent.

\begin{figure}[t] 
\centering
\subfloat[$l=1$]{\includegraphics[scale=0.75]{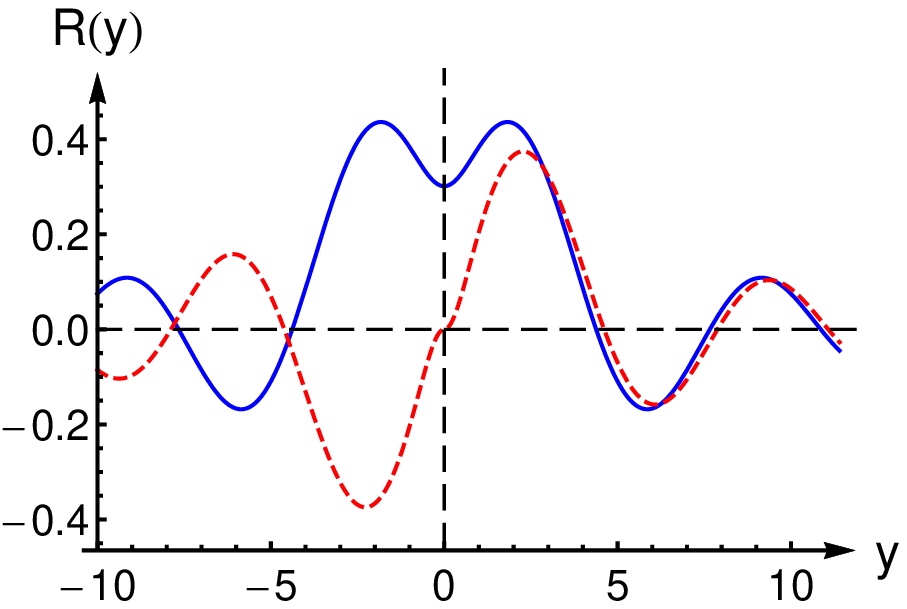}}\hspace{1cm}
\subfloat[$l=3$]{\includegraphics[scale=0.75]{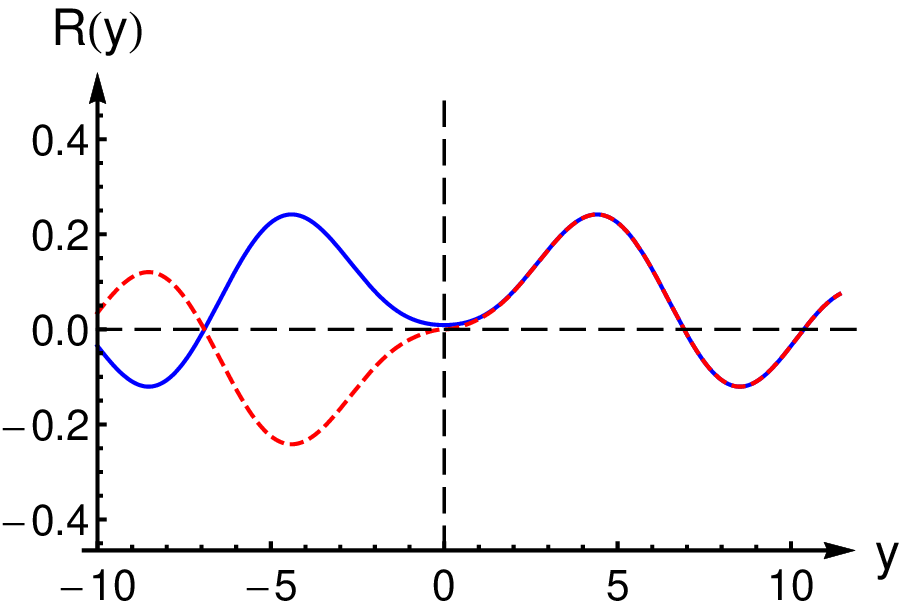}}
\caption{Behavior of the radial functions $R_{kl}(y)$ for $l=1$ and $l=3$,
with $k=1$ and  $b=1$. The blue solid line corresponds to $R_{kl}(y)$
for the manifold  $\mathcal{M}_{b}$ [see Eq. \eqref{eq:rad_sol_MbD_y}],
while the red short-dashed line describes $R_{kl}(y)$
for the manifold $\widehat{M}_{b}$
[see Eq. \eqref{eq:rad_sol_Mb_y_S}]. The two solutions coincide for even values of $l$.
The solutions for $\mathcal{M}_{b}$ have been obtained by imposing that, as $y\rightarrow+\infty$, the functions $R_{kl}(y)$
approach the solutions for Minkowski spacetime [see Eq. \eqref{eq:rad_sol_Min}].}
\label{fig:3}
\end{figure}

The behavior of the solutions $R_{kl}(y)$ obtained for the manifold
$\widehat{M}_{b}$
and for the smooth defect manifold $\mathcal{M}_{b}$ is compared in Fig. \ref{fig:3}.
In App. \ref{app:radial_solution_0}, we provide a detailed derivation
of the radial solutions for $\mathcal{M}_{0}$ and $\mathcal{M}_{b}$ near $y=0$,
which shows how the different behavior of the two solutions originates.
In brief, this is due to the fact that both Minkowski spacetime $M\approx\mathcal{M}_0$ and the defect manifold $\mathcal{M}_b$ are smooth manifolds whose metrics are invariant under parity. This implies that the solutions $\mathbf{\Phi}_{klm}$ of the Klein--Gordon equation for both cases must be regular everywhere and be parity eigenstates. However, spherical coordinates are ill defined at $r=0$ and, consequently, the radial equation \eqref{eq:rad_eq_Min} is singular at $r=0$ [in particular the last term of the left-hand side proportional to $l(l+1)$]. Then, in order for the solution to be regular at $r=0$, it must be $R(r)\sim r^{l}$ for $r\sim0$. For $\mathcal{M}_b$, on the other hand, the point $r=0$ does not belong to the manifold ($r=\sqrt{y^2+b^2}\geq b>0$). It follows that the last term of the left-hand side of the radial equation \eqref{eq:rad_eq_S} is regular everywhere and,  consequently, the parity of the solutions does not depend on the value of $l$. Near $y=0$, the regularity of the $\mathcal{M}_{b}$ solution implies $R(y)\sim y^0$.

\subsection{Scalar solution for the nonsmooth defect}
\label{subsec:nonsmooth_defect_solutions}

The last case we have to discuss is the defect manifold $\widetilde{\mathcal{M}}_{b}$ described by the metric \eqref{eq:nonsmooth-def-metric}.
This metric is not differentiable at the defect surface $\widetilde{y}=0$, which makes it problematic to study the
Klein--Gordon equation near $\widetilde{y}=0$.
In fact, the Klein--Gordon equation is given by Eq. \eqref{eq:klein_gordon_general}, which contains derivatives of the metric and
it is not defined at the defect boundary. This can be seen explicitly examining the radial equation \eqref{eq:rad_eq_NS}, where the coefficient of the first
derivative turns out to be discontinuous at $\widetilde{y}=0$. Consequently, proper solutions of the Klein--Gordon equation can only be found for
$\widetilde{y}\neq0$.

What we can do is to construct a global ``solution''
which solves the radial equation separately in the two regions $\widetilde{y}>0$ and $\widetilde{y}<0$, where the equation is well-defined~\cite{Endnote6}.
Then, we can try to match these solutions at the defect boundary by imposing a continuity condition
\begin{equation}\label{eq:continuity_condition}
\lim_{\widetilde{y}\rightarrow0^+}R_{kl}^{(\widetilde{\mathcal{M}}_{b})(+)}(\widetilde{y})=
\lim_{\widetilde{y}\rightarrow0^-}R_{kl}^{(\widetilde{\mathcal{M}}_{b})(-)}(\widetilde{y})\,,
\end{equation}
where $R_{kl}^{(\widetilde{\mathcal{M}}_{b})(+)}(\widetilde{y})$ is the solution obtained in the region $\widetilde{y}>0$
and $R_{kl}^{(\widetilde{\mathcal{M}}_{b})(-)}(\widetilde{y})$ is the solution obtained in the region $\widetilde{y}<0$.
The global ``solution'' is then given by
\begin{equation}
R_{kl}^{(\widetilde{\mathcal{M}}_{b})}(\widetilde{y})=
\left\{\begin{split}&R_{kl}^{(\widetilde{\mathcal{M}}_{b})(+)}(\widetilde{y})\,,
\hspace{0.5cm}&&\text{for\;\;\;}\widetilde{y}>0\,,\\
&R_{kl}^{(\widetilde{\mathcal{M}}_{b})(-)}(\widetilde{y})\,,
\hspace{0.5cm}&&\text{for\;\;\;}\widetilde{y}<0\,.
\end{split}\right.\end{equation}
Since, in these two regions, $\widetilde{\mathcal{M}}_{b}$ is locally diffeomorphic to the manifolds studied in the previous cases, the solutions
$R_{kl}^{(\widetilde{\mathcal{M}}_{b})(\pm)}(\widetilde{y})$ can be immediately obtained by applying the appropriate change of coordinates to the previous solutions.

Considering the smooth manifold $\mathcal{M}_{b}$
where the radial solution is given by
Eq. \eqref{eq:rad_sol_MbD_y}, applying
the change of coordinates \eqref{eq:transformation_y_y}, and imposing the continuity condition \eqref{eq:continuity_condition} gives the expression
\begin{equation}\label{eq:rad_sol_NS}
R_{kl}^{(\widetilde{\mathcal{M}}_{b})(A)}(\widetilde{y})=j_l(k(b+|\widetilde{y}|))\,,\hspace{1cm}\forall\,\widetilde{y}\neq0\,,
\end{equation}
where the condition $\widetilde{y}\neq0$ emphasizes the fact that this is not a proper solution at $\widetilde{y}=0$.
Observe that this expression, as the one obtained for $\mathcal{M}_{b}$,
transforms as follows under point reflections:
\begin{equation}
R_{kl}^{(\widetilde{\mathcal{M}}_{b})(A)}(\widetilde{y})\overset{P}\longrightarrow R_{kl}^{(\widetilde{\mathcal{M}}_{b})(A)}(-\widetilde{y})=
(+1)\, R_{kl}^{(\widetilde{\mathcal{M}}_{b})(A)}(\widetilde{y})\,,
\end{equation}
with a nonstandard parity eigenvalue of $+1$.

Another acceptable global ``solution'' can be obtained from the solution derived for $\widehat{M}_{b}$. In that case, we have already applied the
appropriate change of coordinates which leads to Eq. \eqref{eq:rad_sol_Mb_y_NS}. Observing that this function is continuous at $\widetilde{y}=0$
and restricting to the real part of this expression, we obtain
\begin{equation}
R_{kl}^{(\widetilde{\mathcal{M}}_{b})(B)}(\widetilde{y})=\operatorname{Re}\left[R_{kl}^{(\widehat{M}_{b})}(\widetilde{y})\right]\,,\hspace{1cm}\forall\,\widetilde{y}\neq0\,,
\end{equation}
where $R_{kl}^{(\widehat{M}_{b})}(\widetilde{y})$ is given by Eq. \eqref{eq:rad_sol_Mb_y_NS} and, again, we emphasize that this is a proper solution
of the radial equation \eqref{eq:rad_eq_NS} only for $\widetilde{y}\neq0$.
Observe that this expression behaves as follows under point reflection:
\begin{equation}
R_{kl}^{(\widetilde{\mathcal{M}}_{b})(B)}(\widetilde{y})\overset{P}\longrightarrow R_{kl}^{(\widetilde{\mathcal{M}}_{b})(B)}(-\widetilde{y})=
(-1)^{l}\, R_{kl}^{(\widetilde{\mathcal{M}}_{b})(B)}(\widetilde{y})\,,
\end{equation}
with standard parity eigenvalues $(-1)^{l}$.

Both expressions $R_{kl}^{(\widetilde{\mathcal{M}}_{b})(A)}(\widetilde{y})$ and $R_{kl}^{(\widetilde{\mathcal{M}}_{b})(B)}(\widetilde{y})$ are acceptable
global ``solutions'' of the radial equation for $\widetilde{\mathcal{M}}_{b}$,
Eq. \eqref{eq:rad_eq_NS},
but they have different behavior under parity.
We conclude that, while for $\widehat{M}_{b}$ and $\mathcal{M}_{b}$ the scalar solutions
have definite parity, for $\widetilde{\mathcal{M}}_{b}$ the parity of the solutions is not determined.

\section{Conclusion}\label{sec:conclusion}

In this article, we have considered 
three different spacetime defects (with length scale $b\ne 0$), whose manifolds 
are homeomorphic but not diffeomorphic, and have compared the solutions of the massless Klein--Gordon equation.
We observe that, when the size of the defect is set to zero ($b=0$), the three manifolds are diffeomorphic to Minkowski spacetime and the scalar solutions
are indeed equivalent to each other.

The main result of this article is that, for $b>0$,
the scalar solutions over the smooth defect manifold $\mathcal{M}_{b}$
from Ref.~\cite{KlinkhamerRahmede2013} have different parity compared
to the solutions for standard Minkowski spacetime $M$ and for the
defect manifold $\widehat{M}_{b}$ from Ref.~\cite{BernadotteKlinkhamer2006}
with boundary conditions \eqref{eq:bound_cond_scalar} on the scalar field.
The solutions for the nonsmooth defect manifold $\widetilde{\mathcal{M}}_{b}$
from Ref.~\cite{Schwarz2010} have no definite parity,
since the Klein--Gordon equation is not defined on the defect surface.
The parity eigenvalues of the different solutions are collected
in Table \ref{tab:Parity}. The different parity eigenvalues for $M$
and $\mathcal{M}_{b}$ illustrate the fact that the
latter manifold is non-Lorentzian
[obeying a weaker variety of the elementary flatness condition
as explained in the paragraph below \eqref{eq:smooth_curvature_scalars}
in Sec.~\ref{subsec:smooth_defect}].

\begin{table}[t]
\renewcommand{\tabcolsep}{0.25pc}   
\renewcommand{\arraystretch}{1.1}   
\centering
\begin{tabular}{l|c}
\text{Manifold}&$P$\\
\hhline{=|=}
$\hspace{0cm}M\hspace{0cm}$&$(-1)^{l}$\\
\hline
$\hspace{0cm}\widehat{M}_{b}+\text{boundary\;conditions}\hspace{0cm}$   
\;\;&$(-1)^{l}$\\
\hline
$\hspace{0cm}\widetilde{\mathcal{M}}_{b}\hspace{0cm}$&\text{Ambiguous}\\  
\hline
$\hspace{0cm}\mathcal{M}_{b}\hspace{0cm}$&$+1$\\
\hline
\end{tabular}
\caption{Parity eigenvalues $P$ of the scalar field solutions
for Minkowski spacetime $M$ and
three defect manifolds with length scale $b \ne 0$.}
\label{tab:Parity}
\end{table}

The heuristic explanation for having only $+1$ parity eigenvalues
in Table~\ref{tab:Parity} for the $\mathcal{M}_{b}$ manifold
is as follows.
The original construction of this manifold~\cite{KlinkhamerRahmede2013}
relies on the combination $\sqrt{b^2+y_{1}^2}$
for the chart-1 coordinate $y_{1}$ (and similarly for the charts 2 and 3).
In turn, the corresponding scalar solution involves
the same combination $\sqrt{b^2+y_{1}^2}$. The explicit radial solution is then
given by Eq. \eqref{eq:rad_sol_MbD_y} and shows a nonvanishing value at $y_{1}=0$,
which rules out having continuous solutions that are odd in $y_{1}$.

From the comparison in Table~\ref{tab:Parity},
we conclude that the three spacetime defects,
even if they are topologically equivalent,
produce different modifications of the propagation of the scalar
field~\cite{Endnote7}.
These different effects become particularly important in the context
of a spacetime foam, where the quantum fluctuations of spacetime
may give rise to a ``gas'' of defects in the emerging classical spacetime.
Depending on which particular type of spacetime
defect turns out to be relevant, different modified dispersion relations result.
This conclusion also holds for photons,
for which analogous results can be derived.

\section*{Acknowledgments}

This work has been supported, in part,
by the German Research Foundation 
(DFG) under Grant No. KL 1103/2-1. We thank
the referee for useful comments and
A. Kern for spotting a few minor errors in the Appendix.

\begin{appendix}

\section{Radial solution near $y=0$}
\label{app:radial_solution_0}

The results of Sec. \ref{sec:scalar_solutions} have shown that the scalar field solutions for Minkowski spacetime
[Eqs. \eqref{eq:scal_sol_0_r} and \eqref{eq:scal_M0_by}]
and those for the smooth defect metric [Eq. \eqref{eq:scal_sol_MbD_y}] are equivalent in the separate regions $y>0$ and $y<0$, but not at the origin $y=0$.
In this appendix, we study, for a neighborhood of the origin,
the radial wave equation of the smooth defect
manifold $\mathcal{M}_{b}$ ($b>0$) and the radial wave equation of Minkowski spacetime $M\approx \mathcal{M}_{0}$,
both expressed in the same coordinate system $\{t,y,z,x\}$.

The two radial equations are
\begin{subequations}\label{eq:radeq2}
\begin{align}
&\partial^2_yR+\frac{2}{y}\partial_yR+\left(k^2-\frac{l(l+1)}{y^2}\right)R=0\,,
&\text{for\;\;\;}\mathcal{M}_{0}\label{eq:radeq2a}\,,
\\[2mm]
&\frac{y^2+b^2}{y^2}\partial^2_yR+\frac{2y^2-b^2}{y^3}\partial_yR
+\left(k^2-\frac{l(l+1)}{y^2+b^2}\right)R=0\,,
&\text{for\;\;\;}\mathcal{M}_{b}\label{eq:radeq2b}\,,
\end{align}
\end{subequations}
where the indices $k$ and $l$  on $R_{kl}$ have been dropped.
Since the origin is a regular singular point for both equations, 
we can use the Frobenius method~\cite{TenembaumPollard1963} to find the solutions
around $y=0$. This method allows us to find solutions of the form
\begin{equation}\label{eq:frob1}
R(y)=y^{s}\sum_{n=0}^{\infty}c_n\,y^n\,,\hspace{1cm}c_0\neq0\,.
\end{equation}
Instead of inserting this expression directly into \eqref{eq:radeq2},
it turns out to be useful to rewrite Eq. \eqref{eq:radeq2} in the following form:
\begin{equation}\label{eq:ode}
y^2\,R''+y\,p(y)R'+q(y)R=0\,,
\end{equation}
and to expand its coefficients $p(y)$ and $q(y)$ in powers of $y$: $p(y)=p_0+p_1\,y+p_2\,y^2+...\,$ and $q(y)=q_0+q_1\,y+q_2\,y^2+...\,$.
Inserting the expansion \eqref{eq:frob1} into Eq. \eqref{eq:ode} we obtain
\begin{equation}\label{eq:ode1}
\sum_{n=0}^{\infty}\left[(n+s)(n+s-1)+p(y)(n+s)+q(y)\right]c_n\,y^{n+s}=0\,.
\end{equation}
For $n=0$, we have
\begin{equation}
\left[s^2+(p_0-1)s+q_0\right]c_0\, y^s=0\,,
\end{equation}
and, assuming $c_0\neq0$, it must be that
\begin{equation}\label{eq:indicial-equation}
s^2+(p_0-1)s+q_0=0\,.
\end{equation}
This last equation is called the ``indicial equation''~\cite{TenembaumPollard1963},  which has,
in general, two roots $(s_1,s_2)$ corresponding to the two independent solutions. We assume that these roots are real and that
$s_1\geq s_2$. Then, one solution is always given by Eq. \eqref{eq:frob1} with $s=s_1$. If the difference of the two
roots $\Delta s=s_1-s_2$ is an integer, the second solution takes the form
\begin{equation}\label{eq:frob2}
R_2(y)=
\alpha\,R_1(y)\,\frac{1}{2}\,\log y^2   
+y^{s_2}\sum_{n=0}^{\infty}d_n\,y^n\,,\hspace{1cm}d_0\neq0\,.
\end{equation}

Inserting the power series for $p(y)$ and $q(y)$,  
Eq. \eqref{eq:ode1} can now be written as 
\begin{equation}\begin{split}\label{eq:ode2}
y^s\sum_{n=0}^{\infty}\Big\{&\left[(n+s)(n+s-1)+(n+s)p_0+q_0\right]c_n
+\left[(n+s-1)p_1+q_1\right]c_{n-1}\\
&+\left[(n+s-2)p_2+q_2\right]c_{n-2}+...\Big\}\, y^n=0\,,
\end{split}\end{equation}
from which we obtain the coefficients $c_n$ of the first solution:
\begin{equation}\label{eq:coeff}
c_n=-\frac{\sum_{i=1}^{\infty}\left[(n+s_1-i)p_i+q_i\right]c_{n-i}}{\left[(n+s_1)(n+s_1-1)+(n+s_1)p_0+q_0\right]}\,.
\end{equation}
In order to obtain the coefficients of the second solution when $\Delta s$ is an integer, we have to insert the expression
\eqref{eq:frob2} into the differential equation \eqref{eq:ode}. After some algebra, we arrive at the following recursion relation:
\begin{align}\label{eq:coeff1}
&d_n= \nonumber \\
&-\frac{\sum\limits_{i=1}^{n}\left[(n+s_2-i)p_i+q_i\right]d_{n-i}
+\alpha\left[(2(n+s_2)-1)c_{n-\Delta s}
+\sum\limits_{i=0}^{n-\Delta s}c_{n-\Delta s-i}
\,p_i\right]}{(n+s_2)(n+s_2-1)+(n+s_2)p_0+q_0}\,,
\end{align}
from which we obtain also the value of $\alpha$.

Returning to Eqs. \eqref{eq:radeq2a} and \eqref{eq:radeq2b},
we observe that the functions
$p(y)$ and $q(y)$ and their expansions are:
\begin{subequations}
\begin{align}
&\left(\begin{aligned}
&p(y)=2\\
&q(y)=-l(l+1)+k^2\,y^2
\end{aligned}\right)\,,&\text{for\;\;\;}\mathcal{M}_{0}\,,\\[4mm]
&\left(\begin{aligned}
&p(y)=\frac{2y^2-b^2}{y^2+b^2}=-1+\frac{3}{b^2}\,y^2-\frac{3}{b^4}\,y^4+...\\[2mm]
&q(y)=\frac{y^4}{y^2+b^2}\left(k^2-\frac{l(l+1)}{y^2+b^2}\right)=
\frac{b^2k^2-l(l+1)}{b^4}\,y^4+...
\end{aligned}\right)\,,&\text{for\;\;\;}\mathcal{M}_{b}\,,
\end{align}\end{subequations}
from which we can read off the coefficients $p_i$ and $q_i$. 
The indicial equation \eqref{eq:indicial-equation} then has the following roots: 
\begin{subequations}
\begin{align}
&s_1=l\,, \quad s_2=-(l+1)\,,
&\text{for\;\;\;}\mathcal{M}_{0}\,,\\[4mm]
&s_1=2\,, \quad s_2=0\,,
&\text{for\;\;\;}\mathcal{M}_{b}\,,
\end{align}\end{subequations}
where $\Delta s$ is an integer in both cases. We observe a crucial difference:
the roots for $\mathcal{M}_{0}$ depend on $l$, whereas
the roots for $\mathcal{M}_{b}$ are independent of $l$.
This different behavior of the roots is the origin of the
different behavior of the solutions under parity transformations.

\begin{figure}[t]  
\centering
\subfloat[$\mathcal{M}_{0}$]{\includegraphics[scale=0.75]{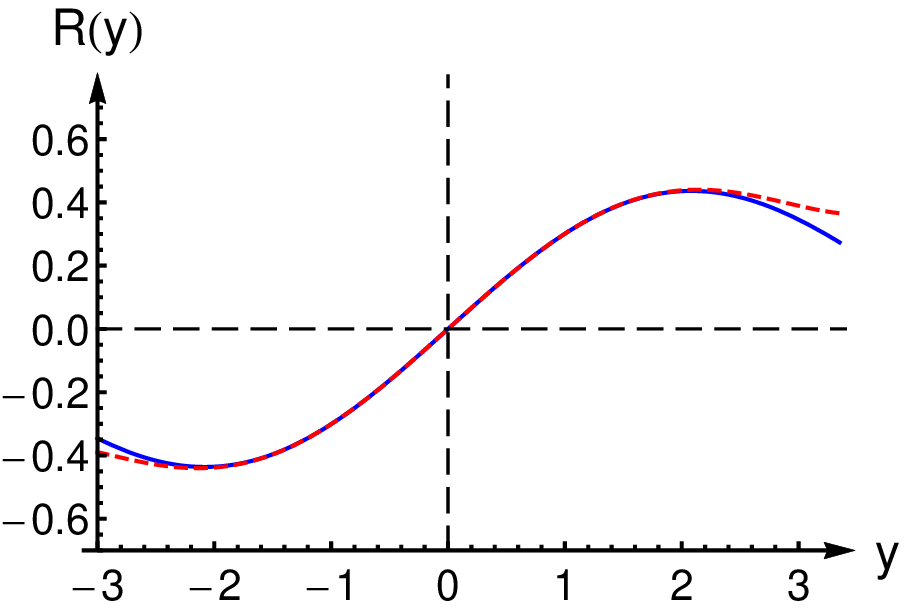}}
\hspace*{5mm}
\subfloat[$\mathcal{M}_{b}$]{\includegraphics[scale=0.75]{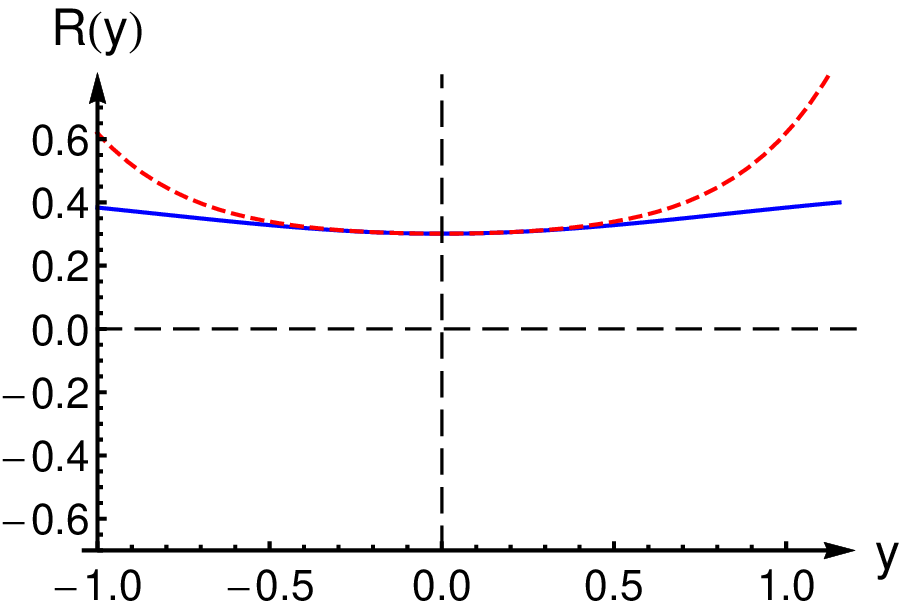}}
\caption{Comparison of truncated radial functions
(red short-dashed lines) derived in Appendix~\ref{app:radial_solution_0}
and exact solutions  (blue solid lines) from 
Secs.~\ref{subsec:minkowski_space} and \ref{subsec:smooth_defect_solutions}  
for $l=1$, $k=1$, and $b=1$.
Figure (a) depicts the results obtained for $\mathcal{M}_{0}$
(which are equivalent to the results obtained in Minkowski spacetime $M$).
Figure (b) depicts the results obtained for the smooth defect manifold $\mathcal{M}_{b}$ with $b>0$.
The red short-dashed lines correspond to the solutions \eqref{eq:frobsolcompl}, where the sums have been truncated at order $y^{4+s}$ ($s$ being the corresponding root of the indicial equation).
The blue solid lines in (a) and (b)
represent the exact radial functions given by, respectively,
Eqs. \eqref{eq:rad_sol_Min_y} and \eqref{eq:rad_sol_MbD_y}.}\label{fig:4}
\end{figure}

Now, the coefficients of the first solution (corresponding to the largest root) are easily obtained from Eq. \eqref{eq:coeff},
where we can set $c_0=1$ without any loss of generality:
\begin{subequations}\label{eq:frobsol1}
\begin{align}
&R_1^{(\mathcal{M}_{0})}(y)=
y^{l}\left\{1-\frac{k^2}{4l+6}\,y^2+\frac{k^4}{8(4l(l+4)+15)}\,y^4+...\right\}\,,
\label{eq:frobsol1a}\\[2mm]
&R_1^{(\mathcal{M}_{b})}(y)=
y^2\left\{1-\frac{3}{4b^2}\,y^2+\frac{l(l+1)-b^2k^2+15}{24b^4}\,y^4+...\right\}\,.
\label{eq:frobsol1b}
\end{align}\end{subequations}
The coefficients of the second solution  
can be obtained from Eq. \eqref{eq:coeff1}, setting $d_0=1$,%
\begin{subequations}\label{eq:frobsol2}
\begin{eqnarray}
R_2^{(\mathcal{M}_{0})}(y) &=&
\alpha R_1^{(\mathcal{M}_0)}(y)\,\frac{1}{2}\,\log y^2 
+\alpha\frac{1-2l}{2l}c_{-2l}\, y+\frac{k^2+\alpha c_{1-2l}(3-2l)}{4l-2}\, y^2
\nonumber \\ &&
+\alpha\frac{k^2 c_{-2l}
(1-2l)/(2l)+c_{2-2l}(5-2l)}{6l-6}\, y^3
\nonumber \\ &&
+
\frac{k^2(k^2+\alpha c_{1-2l}(3-2l))/(4l-2)
+\alpha c_{3-2l}(7-2l)}{8l-12}\, y^4
+\dots\bigg\rbrace\,,
\label{eq:frobsol2a}
\\[2mm]
R_2^{(\mathcal{M}_{b})}(y)&=&
y^0\left\{1+d_2\,y^2+\frac{l(l+1)-b^2(k^2+6d_2)}{8b^4}\,y^4+...\right\}\,.
\label{eq:frobsol2b}
\end{eqnarray}
\end{subequations}
Several comments are in order. The parameter $\alpha$ in $R_2^{(\mathcal{M}_{0})}(y)$ depends on the value of $l$
(for example,  we obtain $\alpha=0$ for $l=1$). The coefficient $d_2$ in $R_2^{(\mathcal{M}_{b})}(y)$ is undetermined, so
that we can choose to set it to zero ($d_2=0$).
The general solutions are, then, given by
\begin{subequations}\label{eq:frobsolcompl}
\begin{align}
R_{kl}^{(\mathcal{M}_{0})}(y)=
\overline{a}\,R_1^{(\mathcal{M}_{0})}(y)
+\overline{b}\,R_2^{(\mathcal{M}_{0})}(y)   
\label{eq:frobsolcompla}\,,\\[2mm]
R_{kl}^{(\mathcal{M}_{b})}(y)=
\overline{c}\,R_1^{(\mathcal{M}_{b})}(y)
+\overline{d}\,R_2^{(\mathcal{M}_{b})}(y)\,.
\label{eq:frobsolcomplb}
\end{align}\end{subequations}
We observe that $R_1^{(\mathcal{M}_{0})}(y)$ is proportional to $j_l(ky)$
and, as expected, has parity $(-1)^{l}$
[the second solution $R_2^{(\mathcal{M}_{0})}(y)$ is proportional to $y_l(ky)$].
For $\mathcal{M}_{b}$, both solutions turn out to have parity $+1$ and it is impossible to build a
solution that is odd under parity.

In Fig.~\ref{fig:4}, we show the behavior of the truncated $l=1$
functions from Eq. \eqref{eq:frobsolcompl}
compared to the exact solutions,
Eqs. \eqref{eq:rad_sol_Min_y} and \eqref{eq:rad_sol_MbD_y}.
The coefficients $\overline{a}$, $\overline{b}$, $\overline{c}$, and $\overline{d}$ 
are obtained by imposing that, at the origin,  the solutions \eqref{eq:frobsolcompl}
and their derivatives coincide with the exact solutions:
\begin{subequations}
\begin{align}
R_{kl}^{(\mathcal{M}_{0})}(y)&
=j_l(ky)
= \alpha_{l}\, (ky)^{l} + \alpha_{l+2}\, (ky)^{l+2} + \cdots \,, 
\hspace{1cm}&\text{for\;\;\;}y\sim0 \,,
\\[2mm]
R_{kl}^{(\mathcal{M}_{b})}(y)&
=j_l(k\sqrt{y^2+b^2})
= \beta_{0} + \beta_{2} \,(ky)^{2} + \cdots  \,,  
\hspace{1cm}&\text{for\;\;\;}y\sim0 \,.
\end{align}\end{subequations}
The different behavior at $y=0$ of the solutions
over Minkowski spacetime $M$ (Fig.~\ref{fig:4}--a)
and the smooth defect manifold $\mathcal{M}_{b}$
(Fig.~\ref{fig:4}--b) is manifest.

\end{appendix}



\begin{thebibliography}{99}
\bibitem{Endnote1}
These results are obtained in the sense of distributions.
The metric components are not differentiable at $\widetilde{y}=0$.
\bibitem{Endnote2}
When taking the limit $b\rightarrow0$, we must remember
to add a point to the manifold, in order to fill-in the point-like remnant
of the spacetime defect at the origin: $\widetilde{\mathcal{M}}_{0}=
\lim_{b\rightarrow0}\widetilde{\mathcal{M}}_{b}+\{0\}$.
\bibitem{Endnote3}
The change of coordinates
\eqref{eq:non_smooth_rad_change_of_coord} 
is still nondifferentiable at $\widetilde{y}=0$.  
However, this point coincides with the origin of the spherical coordinate
system, which itself is ill defined (the atlas obtained by spherical coordinates does not include the origin).
\bibitem{Endnote4}
The factors $(y^2+b^2)^{1/2}$ in the metrics from \eqref{eq:smooth_mass_def_metric}
or \eqref{eq:smooth_def_metric} can be replaced
by other smooth combinations of $y$ and $b$, for example,
$(y^{2n}+b^{2n})^{1/(2n)}$ with integer $n\geq 2$.
This freedom is a manifestation of the general coordinate invariance.
\bibitem{Endnote5}
Hence, $y$ is a type of ``pre-radial'' coordinate
with the sign of $y$ corresponding to the different sides of the
defect.
\bibitem{Endnote6}
The quotation marks emphasize that this is not a proper solution as defined in Sec. \ref{subsec:General-solution}.
\end{thebibliography}
\end{document}